\documentclass[aps,preprint,amsmath,amssymb]{revtex4}
\usepackage{graphicx}
\begin{document}

\title{Exclusive excited leptons search in two lepton
final states at the CERN-LHC}

\author{S.C. \.{I}nan}
\email[]{sceminan@cumhuriyet.edu.tr}
\affiliation{Department of Physics, Cumhuriyet University,
58140, Sivas, Turkey}

\begin{abstract}
We study the potential of the exclusive $pp\to p\ell^-\ell^+p$ process to probe excited leptons at the LHC which is known to be one of the most clean channel at the hadron colliders. The sensitivity of the model parameters is obtained 95\% confidence level by considering three forward detector acceptances; $0.0015< \xi < 0.5$, $0.0015< \xi <0.15$ and $0.1< \xi < 0.5$.
\end{abstract}


\maketitle

\section{Introduction}
The Standard Model (SM) left some questions unanswered such
as the number of fermion generation and fermion mass spectrum.
Attractive solutions are provided by models assuming compositeness.
A typical consequence of compositeness is the assumption of excited leptons
and quarks. In this model, charged and neutral leptons should be regarded as the composite
structures which made up of more fundamental constituents. The existence of such quark and leptons substructure should be regarded as the ground state of a rich spectrum of new particles with new quantum numbers. Therefore, the discovery
of excited quarks and leptons would be different signal for compositeness. All composite models have an underlying
substructure which is characterized by a scale $\Lambda$. For the precise measurement of electron
and muon g-2 and theoretical calculations the $\Lambda<10$ $TeV$ is expected \cite{renard, del, suz}.

The Lagrangian describing the transition between SM and excited leptons should conform to the chiral
symmetry so that the  light leptons can not acquire radiatively a large anomalous magnetic moment \cite{bro}. Interactions between excited leptons ($\ell^{*}$) and SM fermions ($\ell$) may be mediated by gauge bosons, described by the $SU(2)\times U(1)$ invariant effective Lagrangian as \cite{cab, kuhn, hagi, bo1, bo2},

\begin{eqnarray}
L=\frac{1}{2\Lambda}\overline{\ell}^{*}\sigma^{\mu\nu}(gf\frac{\overrightarrow{\tau}}{2}\overrightarrow{W}_{\mu\nu}+g^{'}f^{'}\frac{Y}{2}B_{\mu\nu})\ell_{L}+h.c.,
\end{eqnarray}
where the $\overrightarrow{W}_{\mu\nu}$ and $B_{\mu\nu}$ are the field strength tensors of the $SU(2)$ and $U(1)$. $\overrightarrow{\tau}$ and $Y$ are the generators of the corresponding gauge group, $g$ and $g^{'}$ are $SU(2)_{L}$ and $U(1)_{Y}$ coupling constant. The constants $f$ and $f^{'}$ are coupling parameters associated with the three SM gauge groups and are determined by yet unknown composite dynamics. In the physical basis, the Lagrangian can be written as

\begin{eqnarray}
\label{lang}
L=\frac{g_{e}}{2\Lambda}(f-f^{'})N_{\mu\nu}\sum_{\ell=\nu_{e},e}\overline{\ell}^{*}\sigma^{\mu\nu}\ell_{L}+\frac{g_{e}}{2\Lambda}
f\sum_{\ell,\ell^{'}=\nu_{e},e}\Theta^{\overline{\ell}^{*},\ell}_{\mu\nu}\overline{\ell}^{*}\sigma^{\mu\nu}\ell_{L}^{'}+h.c..
\end{eqnarray}
First term in the equations (\ref{lang}) is a purely diagonal term with $N_{\mu\nu}=\partial_{\mu}A_{\nu}-\tan\theta_{W}\partial_{\mu}Z_{\nu}$
and second term is a non-Abelien part which involves triple as well as quartic vertices with

\begin{eqnarray}
\Theta^{\overline{\nu}_e^{*},\nu_{e}}_{\mu\nu}=\frac{2}{\sin2\theta_{W}}\partial_{\mu}Z_{\nu}-i\frac{g_{e}}{\sin^{2}\theta_{W}}W^{+}_{\mu}W^{-}_{\nu},
\end{eqnarray}

\begin{eqnarray}
\Theta^{\overline{e}^{*},e}_{\mu\nu}=-(2\partial_{\mu}A_{\nu}+2\cot2\theta_{W}\partial_{\mu}Z_{\nu}-i\frac{g_{e}}{\sin^{2}\theta_{W}}W^{+}_{\mu}W^{-}_{\nu})
\end{eqnarray}

\begin{eqnarray}
\Theta^{\overline{\nu}_{e}^{*},e}_{\mu\nu}=\frac{\sqrt{2}}{\sin\theta_{W}}(\partial_{\mu}W^{+}_{\nu}-ig_{e}W^{+}_{\mu}(A_{\nu}+\cot\theta_{W}Z_{\nu})),
\end{eqnarray}

\begin{eqnarray}
\Theta^{\overline{e}^{*},\nu_{e}}_{\mu\nu}=\frac{\sqrt{2}}{\sin\theta_{W}}(\partial_{\mu}W^{-}_{\nu}+ig_{e}W^{-}_{\mu}(A_{\nu}+\cot\theta_{W}Z_{\nu})).
\end{eqnarray}

From equation (\ref{lang}), the chiral $V\ell^{*}\ell$ ($V=\gamma, Z, W$) vertex can be obtained as follows

 \begin{eqnarray}
 \label{ver}
 \Gamma^{V\overline{\ell^{*}}\ell}_{\mu}=\frac{g_{e}}{2\Lambda}q^{\nu}\sigma_{\mu\nu}(1-\gamma_{5})f_{V}.
 \end{eqnarray}
Here $q$ is the gauge boson momentum, $f_V$ is the electroweak coupling parameter and defined for photon by $f_{\gamma}=Q_{f}f^{'}+I_{3L}(f-f^{'})$
where we set $f=f^{'}$.

Up to now searches have not found any signal for the excited leptons at the  HERA $ep$ \cite{hera1,hera}, LEP $e^{+}e^{-}$ \cite{lep1,lep2,lep3,lep4} and the Tevatron $p\bar{p}$ \cite{tevatron} colliders. However, in these experimental studies have been found  different limits for mass of the excited lepton ($m_*$) with different $f_\gamma/\Lambda$ and exclusion region for $f_\gamma/\Lambda$-$m_*$ parameter plane. Excited leptons have been also studied LHC \cite{lhc} and next $e^{-}e^{+}$, $e\gamma$ colliders \cite{ebo, cak1, cak2, cak3}.

In this paper, we examine the potential of exclusive $pp \rightarrow p\ell^{-}\ell^{+}p$ reaction
at the LHC to probe excited leptons. The existence of excited leptons would
therefore results in deviations from SM cross-sections. The exclusive production $pp \rightarrow pXp$
has a very clean experimental signature due to absence of the proton remnants. So very clean final states
can be well defined. Intact scattered protons which are scattered at very small angels
after the collisions are detected by very forward detectors (VFD). ATLAS and CMS collaborations
have a program of forward physics with located in a region nearly $100-400$ m from the central dedectors.
VFD's will be constructed as close as a few mm to the beamline. Serious backgrounds can be eliminated by
use of these forward detectors for the following reason. If the energy of the outgoing intact protons is measured by the
VFD's, exclusive or photon-induced process provide an additional method to calculate the invariant mass
of the central system. Photon-induced reactions are electromagnetic in nature which is one of the
advantages of the this type interactions. Moreover, the lepton pair production by two photon fusion
is currently being planned to be used as a luminosity monitor for the LHC \cite{khoze}. The main contamination to these clean environment
comes from the proton dissociation in to the X, Y system, $pp \rightarrow X\ell^{-}\ell^{+}Y$ where $X$ and $Y$
are baryon excitations. This background can be reduced by imposing a cut on the transverse momentum of the photon pair \cite{khoze}.
With this cut lepton pair production is one of the most clean channels at the LHC.

\section{Photon-photon physics at the LHC}

The equivalent photon approximation (EPA) has been successfully applied to describe processes involving photon exchange with proton
beams at the LHC \cite{budnev,baur}. In EPA almost real photons with low virtuality $(Q^{2}=-q^{2})$ are emitted by the incoming
protons producing an object $X$ through the process $pp \rightarrow p\gamma\gamma p \rightarrow pXp$. The photon spectrum
of virtuality $Q^{2}$ and energy $E_{\gamma}$ then become,

\begin{eqnarray}
f=\frac{dN}{dE_{\gamma}dQ^{2}}=\frac{\alpha}{\pi}\frac{1}{E_{\gamma}Q^{2}}
[(1-\frac{E_{\gamma}}{E})
(1-\frac{Q^{2}_{min}}{Q^{2}})F_{E}+\frac{E^{2}_{\gamma}}{2E^{2}}F_{M}].
\label{phs}
\end{eqnarray}
The terms in equation (\ref{phs}) are the following. $E$: energy of the incoming proton, $m_{p}$: proton mass, $E_{\gamma}$: photon energy related to $E$ by $E_{\gamma}=\xi E$, where $\xi$ is the momentum fraction loss of the protons $\xi=(|\vec{p}|-|\vec{p}^{\,\,\prime}|)/|\vec{p}|$. The remaning terms are given as,

\begin{eqnarray}
Q^{2}_{min}=\frac{m^{2}_{p}E^{2}_{\gamma}}{E(E-E_{\gamma})},
\;\;\;\; F_{E}=\frac{4m^{2}_{p}G^{2}_{E}+Q^{2}G^{2}_{M}}
{4m^{2}_{p}+Q^{2}} \\
G^{2}_{E}=\frac{G^{2}_{M}}{\mu^{2}_{p}}=(1+\frac{Q^{2}}{Q^{2}_{0}})^{-4},
\;\;\; F_{M}=G^{2}_{M}, \;\;\; Q^{2}_{0}=0.71 \mbox{GeV}^{2}.
\end{eqnarray}
The magnetic moment of the proton is taken to be $\mu_{p}^{2}=7.78$. $F_{E}$ and $F_{M}$ are functions of the electric and magnetic
form factors. The luminosity spectrum of photon-photon collisions $\frac{dL^{\gamma\gamma}}{dW}$ can be introduced in EPA as,

\begin{eqnarray}
\label{efflum}
\frac{dL^{\gamma\gamma}}{dW}=\int_{Q^{2}_{1,min}}^{Q^{2}_{max}}
{dQ^{2}_{1}}\int_{Q^{2}_{2,min}}^{Q^{2}_{max}}{dQ^{2}_{2}} \int_{y_{
min}}^{y_{max}} {dy \frac{W}{2y} f_{1}(\frac{W^{2}}{4y}, Q^{2}_{1})
f_{2}(y,Q^{2}_{2})}
\end{eqnarray}
where W is the invariant mass of the two photon system $W=2E\sqrt{\xi_{1} \xi_{2}}$ with

\begin{eqnarray}
y_{min}=\mbox{MAX}(W^{2}/(4\xi_{max}E), \xi_{min}E), \;\;\;
y_{max}=\xi_{max}E.
\end{eqnarray}
$Q_{max}^2$ is taken to be $2\,\text{GeV}^2$. Using the equation (\ref{efflum}) the cross section
for the complete process $pp \rightarrow p\gamma\gamma p \rightarrow pXp$ can be obtained as follows,

\begin{eqnarray}
\label{completeprocess}
 d\sigma=\int{\frac{dL^{\gamma\gamma}}{dW}}
\,d\hat {{\sigma}}_{\gamma\gamma \to X}(W)\,dW.
\end{eqnarray}
Here $d\hat {{\sigma}}_{\gamma\gamma \to X}(W)$ cross section of the subprocess $\gamma\gamma \to X$. This exclusive
production phenomenon recently was observed in the measurements of CDF collaborations \cite{cdf1,cdf2,cdf3,cdf4,cdf5,cdf6,cdf7}.
LHC produce high energetic proton-proton collisions with high luminosity. So exclusive production potential at the LHC is important, and
it has being studied some workers \cite{lhc1,lhc2,lhc3,lhc4,lhc5,lhc6,albrow,lhc7,lhc8}.

ATLAS and CMS have central detectors with a psedorapidity $|\eta|$ coverage $2.5$ for the tracking system at the LHC and the $X$ object
through the process $pp \rightarrow p\gamma\gamma p \rightarrow pXp$ , is  detected by the this central detectors. ATLAS and CMS
collaborations have a program of forward physics with extra detectors for a deeper understanding of the physics from very forward region.
ATLAS forward detectors proposed by the ATLAS Forward Physics Collaboration (AFP) which is positioned at $220$ m and $420$ m from the
interaction point with an acceptance $0.0015 < \xi <0.15$ \cite{albrow}. Similarly CMS-TOTEM forward detectors acceptance spans $0.0015 < \xi< 0.5$
and $0.1< \xi <0.5$ \cite{lhc6, avati}. When the forward detector installed closer to the interaction point, higher values of $\xi$ is obtained.

\begin{figure}
\includegraphics{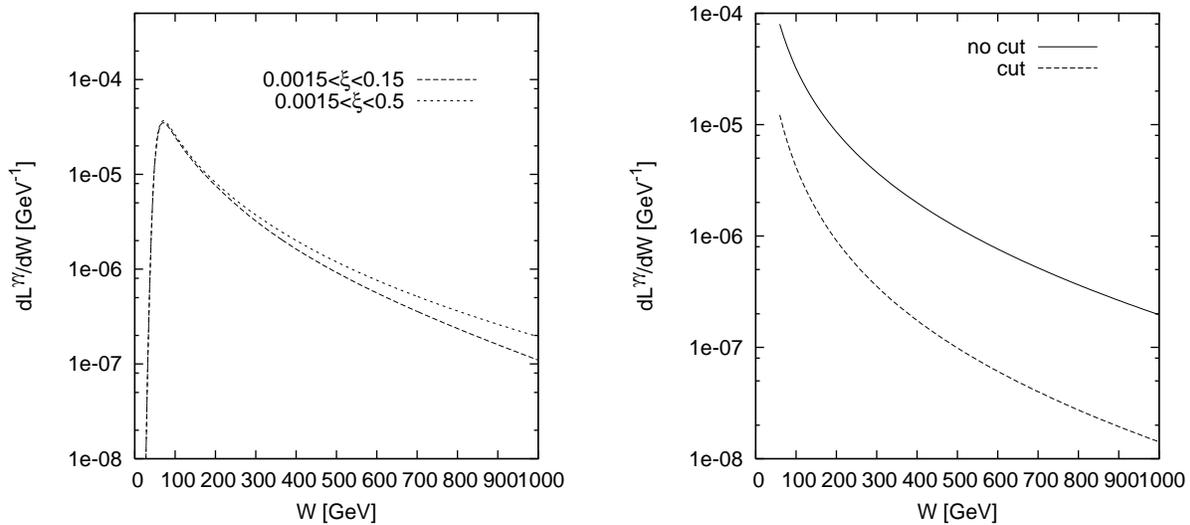}
\caption{Effective $\gamma\gamma$ luminosity as a function of the
invariant mass of the two photon system. Figure on the left shows
effective luminosity for forward detector acceptances
$0.0015<\xi<0.15$ and $0.0015<\xi<0.5$. Figure on the right
represents the cases with and without a cut on transverse momentum
of the photon pair $|\vec{q}_{1t}+\vec{q}_{2t}|<30$MeV. In the right
panel, we do not consider any acceptance i.e., $\xi$ is taken to be
in the interval $0<\xi<1-m_p/E$ where $m_p$ is the mass and $E$ is
the energy of the incoming proton. \label{fig1}}
\end{figure}

As mentioned before the lepton pair production can provide an excellent control sample and at the LHC, this events may also be used to luminosity
monitoring \cite{khoze}. As it was discussed in Ref. \cite{khoze} the main possible contamination comes from the proton dissociation in to
the $X$, $Y$ system such as $pp \to X\ell^{-}\ell^{+}Y$ where $X$ and $Y$ baryon excitations (e.g $N^{*}$, $\triangle$ isobars). To eliminate this background, it was proposed to impose $|\vec{q}_{1t}+\vec{q}_{2t}| <30$MeV cut, where $|\vec{q}_{1t}+\vec{q}_{2t}|$ is vectorial sum of the transverse momentum of the photon pair. In the actual experiment, this cut can be applied on the lepton or photon pair. In our theoretical calculations, we have applied this cut on the photon pair. This is why we have kept integration over $Q^{2}$ in the equation (\ref{efflum}). In figure (\ref{fig1}) in the left side panel, we plot effective $\gamma\gamma$ luminosity as a function of invariant mass of the two photon system for various forward detector acceptances, in the right side panel is the same as left side but for with and without $|\vec{q}_{1t}+\vec{q}_{2t}| <30$ $MeV$ cut. In all the results presented in this work, we impose cut of $|\vec{q}_{1t}+\vec{q}_{2t}| <30$ $MeV$ on transverse momentum of the photon pair.

\section{Phenomenological Results}

The subprocess $\gamma\gamma \to \ell^{+}\ell^{-}$ consists of $t$ and $u$-channel SM  diagrams. New physics contribution
comes from $t$ and $u$-channel excited lepton exchange. The whole polarization summed amplitude has been calculated
in terms of Mandelsam invariants $\hat{s}$, $\hat{t}$ and $\hat{u}$ as,

\begin{eqnarray}
\label{amp}
|M|^{2}=\frac{4g_{e}^{4}f_{\gamma}^4 m_*^{2}}{\Lambda^{4}}( \frac{\hat{s}\hat{t}^{2}}{(\hat{t}-m_*^{2})^{2}}+\frac{\hat{s}\hat{u}^{2}}{(\hat{u}-m_*^{2})^{2}}
+\frac{2\hat{s}\hat{t}\hat{u}}{(\hat{t}-m_*^{2})(\hat{u}-m_*^{2})} )+8g_{e}^{4}(\frac{\hat{u}}{\hat{t}}+\frac{\hat{t}}{\hat{u}})
\end{eqnarray}
where $g_{e}=\sqrt{4\pi\alpha}$ and we omit the mass of the SM leptons. There are no terms of order $(f_{\gamma}/\Lambda)^2$ in equation
(\ref{amp}) since the chiral conserving coupling ensures that the excited lepton diagrams do not interfere with the SM diagrams \cite{vac}. We will
investigate the phenomenology of excited leptons with three forward detector acceptance in next subsections; $0.0015< \xi <0.5$, $0.0015< \xi <0.15$
and $0.1< \xi <0.5$.

\subsection{ CMS - TOTEM with $0.0015< \xi <0.5$ }

First we show the transverse momentum of the final state leptons ($p_{t}$) distribution both SM and total differential cross section
in figure (\ref{fig2}) for $m_*=500$ $GeV$ and $f_{\gamma}/\Lambda=2$ $TeV^{-1}$. It turns out that excited leptons exchange is the dominant effect in high $p_{t}$ scattering. In this motivation
in figure (\ref{fig3}a), we plot cross section of $pp\to p\ell^-\ell^+p$ with and without excited lepton exchange as a function of the minimum transverse momentum ($p_{t, min}$ or $p_{t}$ cut) of the final state leptons for $m_*=500$ $GeV$ and $f_{\gamma}/\Lambda=2$ $TeV^{-1}$. Obviously larger cross sections are obtained with similar curves when the $f_{\gamma}/\Lambda$ values increase. For instance for $f_{\gamma}/\Lambda=4$ $TeV^{-1}$ and $m_*=500$ $GeV$ cross sections are; $0.19$ $fb$ for $p_{t, min}=300$ $GeV$  and $0.086$ $fb$ for $p_{t, min}=500$ $GeV$. We see from figure (\ref{fig3}a) that the deviations from the SM gets to be higher as the $p_t$ cut increases since SM contributions highly peaked in the forward and backward directions due to $\hat{t}$, $\hat{u}=0$ poles as we observe from equations ($\ref{amp}$). Therefore both angular distribution or the $p_t$ cut can be used to improve sensitivity bounds. Hence, we impose different cuts
on the transverse momentum of the final state leptons to decrease the SM events. We should note that Poisson distribution is appropriate sensitivity analysis, since the number of SM events with this cuts is small enough. In this statical analysis, the number of observed events are assumed to be equal to the SM prediction $N_{obs}=L\sigma_{SM}=N_{SM}$ for an integrated luminosity $L$. We impose different $p_t$ cuts to
achieve different $N_{SM}$. In table (\ref{tab1}) we show 95\% C.L. lower bounds on the $m_*$ with different $p_t$ cut and $N_{SM}$ for integrated LHC luminosity values assuming that $f_\gamma/\Lambda=1/m_*$. These bounds are more stringent than the present colliders LEP, Tevatron and HERA \cite{hera1,hera, lep1,lep2, lep3,lep4, tevatron}. Figure (\ref{fig4}a) shows that the 95\% C.L. excluded region in the $f_\gamma/\Lambda-m_*$ plane
for most stringent case $N_{SM}=2$ for the integrated LHC luminosities: $200$ $fb^{-1}$, $100$ $fb^{-1}$ and $50$ $fb^{-1}$. It should be noted that $p_t$ cut not same for different luminosity values to achieve this $N_{SM}$. In figure (\ref{fig4}a) we set cuts of $p_t > 340$ $GeV$ for $L > 200$ $fb^{-1}$, $p_t > 290$ $GeV$ for $L=100$ $fb^{-1}$ and $p_t >240$ $GeV $ for $L = 50$ $fb^{-1}$. Calculations concerning the exclusion regions for other $N_{SM}$ values were done. These results show almost the same curves with convenient transverse momentum cuts on the individual final leptons. We also apply $|\eta|<2.5$ for central detector capacitance. As we see from the figure, excluded area from the process  $pp\to p\ell^-\ell^+p$ extends wider regions than the cases of the colliders LEP, Tevatron and HERA \cite{hera1,hera,lep1,lep2,lep3,lep4,tevatron}.

\subsection{ AFP - CMS with $0.0015< \xi <0.15$}

\begin{figure}
\includegraphics{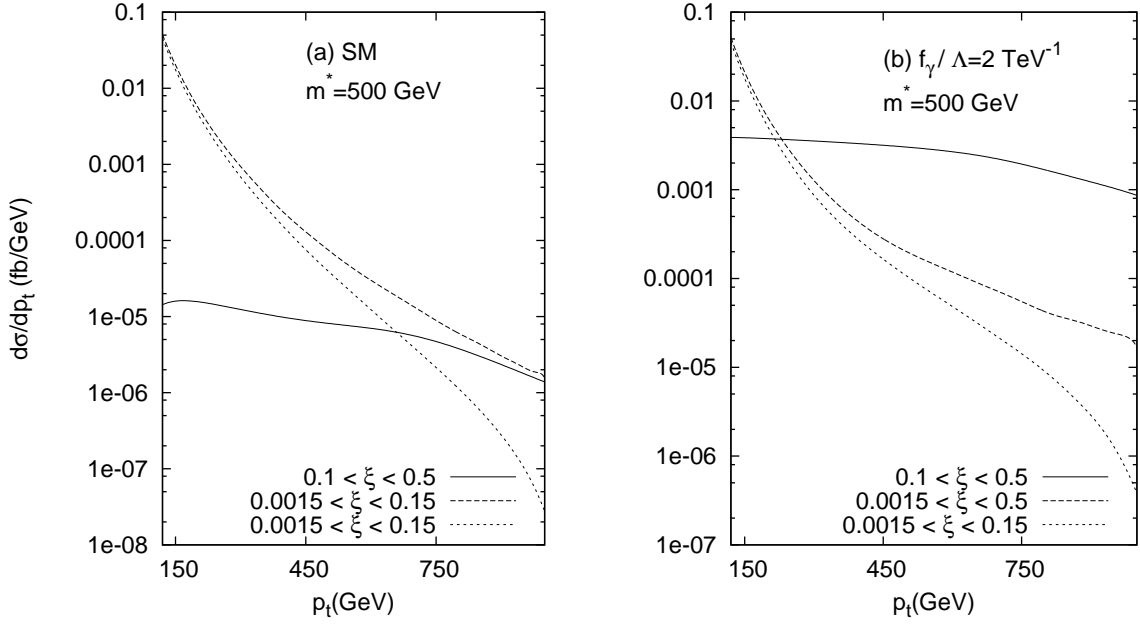}
\caption{Transverse momentum distribution of final state leptons for $pp\to p\ell^-\ell^+p$ for three forward dedector acceptances; $0.0015< \xi <0.5$, $0.0015< \xi <0.5$
and $0.1< \xi <0.5$. Panel (a) is for the SM and Panel (b) is for the $f_{\gamma}/\Lambda=2$ $TeV^{-1}$. $m_*$ is taken to be $500$ $GeV$ and we impose the cuts; $|\vec{q}_{1t}+\vec{q}_{2t}|<30$ $MeV$, $|\eta|<2.5$.\label{fig2}}
\end{figure}

\begin{figure}
\includegraphics{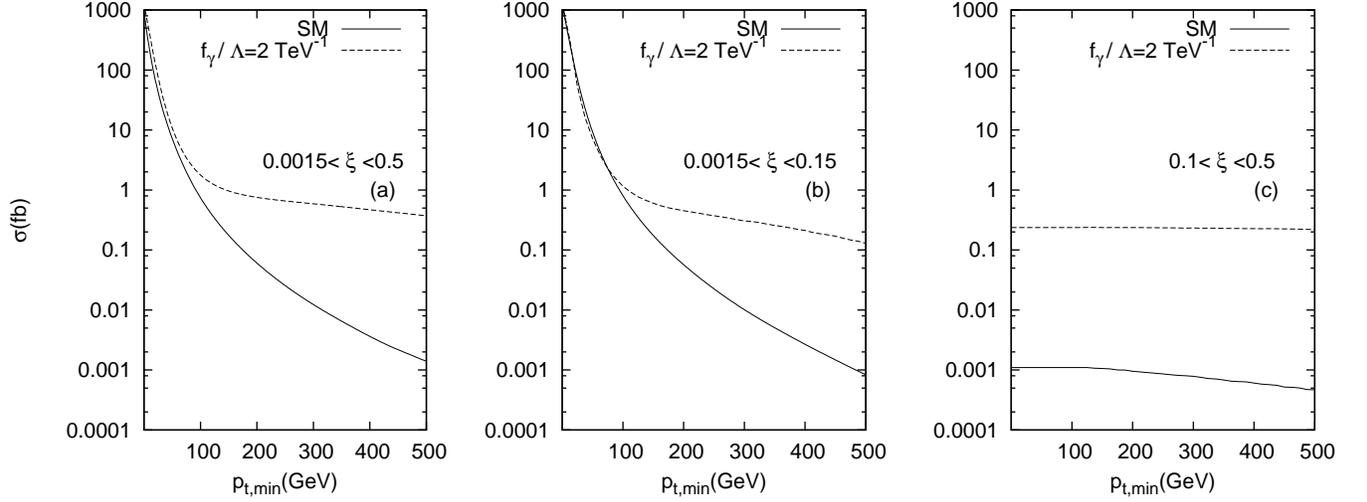}
\caption{Cross section of $pp\to p\ell^-\ell^+p$ as a function of
the transverse momentum cut on the final leptons for three forward dedector acceptances; $0.0015< \xi <0.5$, $0.0015< \xi <0.15$
and $0.1< \xi <0.5$. Solid lines are for the SM and the dotted lines $f_{\gamma}/\Lambda=2$ $TeV^{-1}$.
$m_*$ is taken to be $500$ $GeV$ and we impose the cuts; $|\vec{q}_{1t}+\vec{q}_{2t}|<30$ $MeV$, $|\eta|<2.5$.\label{fig3}}
\end{figure}

\begin{figure}
\includegraphics{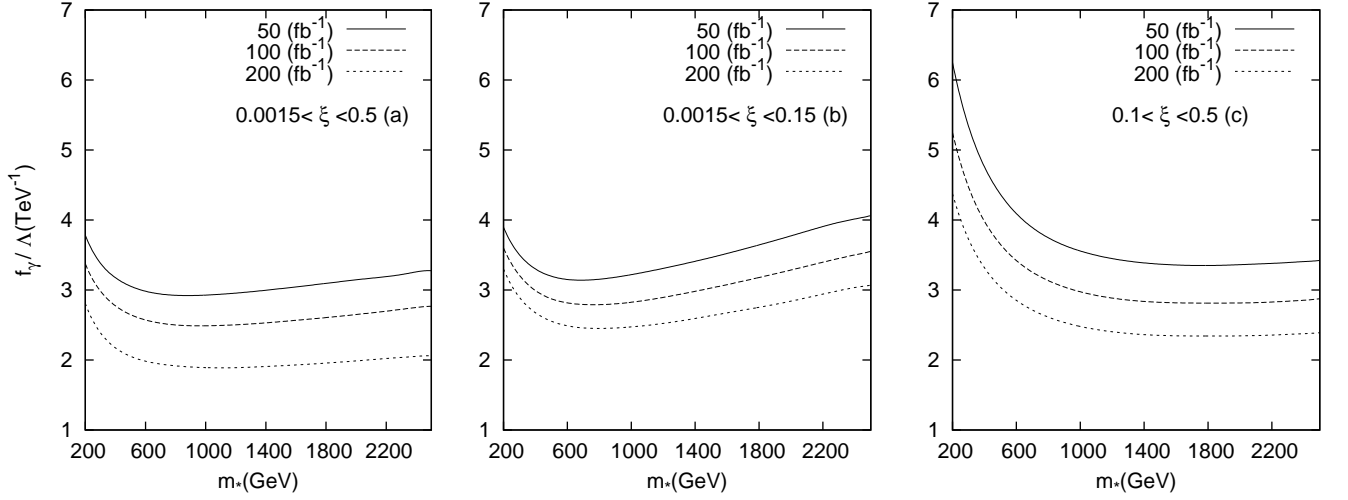}
\caption{95\% C.L. constraints on the parameter plane $f_{\gamma}/\Lambda$ and $m_*$ of $pp\to p\ell^-\ell^+p$ at the three different integrated LHC luminosities; $50 fb^{-1}$, $100 fb^{-1}$, $200 fb^{-1}$ and for three forward dedector acceptances; $0.0015< \xi <0.5$, $0.0015< \xi <0.15$, $0.1< \xi <0.5$. We impose the following cuts: In panel (a); $p_t > 340$ $GeV$ for $L > 200$ $fb^{-1}$, $p_t > 290$ $GeV$ for $L=100$ $fb^{-1}$, $p_t >240$ $GeV $ for $L=50$ $fb^{-1}$ to achieve $N_{SM}=2$. In panel (b);  $p_t > 320$ $GeV$ for $L = 200$ $fb^{-1}$, $p_t > 275$ $GeV$ for $L=100$ $fb^{-1}$, $p_t >230$ $GeV $ for $L=50$ $fb^{-1}$  to achieve $N_{SM}=1$. In panel (c); $p_t > 10$ $GeV$ for $L = 200$ $fb^{-1}$, $L=100$ $fb^{-1}$ and $L=50$ $fb^{-1}$ to achieve $N_{SM}=0$. Also we impose the cuts $|\vec{q}_{1t}+\vec{q}_{2t}|<30$ $MeV$, $|\eta|<2.5$.\label{fig4}}.
\end{figure}

  Figure (\ref{fig2}b) shows the $p_t$ distribution of the final state lepton at the corresponding forward detector acceptance for $m_*=500$ $GeV$ and $f_{\gamma}/\Lambda=2$ $TeV^{-1}$. We see from the figure (\ref{fig2}b) $0.0015< \xi <0.5$ and $0.0015< \xi <0.15$ cases that they have similar curves. Excited leptons exchange is the dominant effect in high $p_t$ scattering. The cross section of  $pp\to p\ell^-\ell^+p$ is given in figure (\ref{fig3}b) as a function of the $p_t$ cut for same parameters in figure (\ref{fig3}a). As seen in this figure, the deviation from the SM gets to be higher as $p_t$ cut increases. Excited lepton contributions and the SM are well separated from each other for large values of the $p_t$ cut. This behavior is the same as $0.0015< \xi <0.5$
  case. Thus we can apply the procedure used in the previous subsection. By assuming that $f_\gamma/\Lambda=1/m_*$ the 95\% C.L. lower bounds on the $m_*$ with different $N_{SM}$ and $p_t$ cut are given in table (\ref{tab2}) for $200$ $fb^{-1}$, $100$ $fb^{-1}$ and $50$ $fb^{-1}$ integrated LHC luminosity values. The obtained results for $L=200$ $fb^{-1}$ and $L=100$ $fb^{-1}$ are better than the cases of the colliders LEP, Tevatron and HERA but for $L=50$ $fb^{-1}$
results have been excluded by the HERA experiment \cite{hera}. Figure (\ref{fig4}b) is shown that the 95\% C.L. excluded region in the $f_\gamma/\Lambda-m_*$ plane in case of $N_{SM}=1$ which is the most stringent case for $0.0015< \xi <0.15$.  In figure (\ref{fig4}b) we impose the cuts; $p_t > 320$ $GeV$ for $L > 200$ $fb^{-1}$, $p_t > 275$ $GeV$ for $L=100$ $fb^{-1}$, $p_t >230$ $GeV $ for $L=50$ $fb^{-1}$  to achieve $N_{SM}=1$ . Similar curves can be obtained for other $N_{SM}$ values with convenient transverse momentum cuts. We also apply $|\eta|<2.5$ for central detector capacitance. As we see from these figures, excluded area from the process $pp\to p\ell^-\ell^+p$ extends wider region than the cases of the colliders LEP, HERA and Tevatron \cite{hera1,hera,lep1,lep2,lep3,lep4,tevatron}.

\subsection{ CMS - TOTEM with $0.1< \xi <0.5$}

In this forward dedector acceptance region, there is no need to impose high cuts on the transverse momentum of the final state leptons to suppress the SM events number. In this acceptance region, invariant mass of the final leptons is greater than $1400$ $GeV$ due to high lower bound of $\xi$. The SM cross section is very small and therefore it does not need to set a high $p_t$ cut. As shown in figure (\ref{fig2}b)
 differential cross section very small at the all of $p_t$ regions for $m_*=500$ $GeV$ and $f_{\gamma}/\Lambda=2$ $TeV^{-1}$. Also figure ($\ref{fig3}c$) shows that SM events number is smaller then $0.5$ for a luminosity of $200$ $fb^{-1}$ at the all of $p_t$ cuts for same parameters in figure (\ref{fig2}). We have applied only $p_t>10$ $GeV$ and $|\eta|<2.5$ for central detector capacitance. Therefore, figure (\ref{fig3}c) shows that when the $p_t$ cut increases, total cross section is not affected. We have obtained 95\% C.L. constraints on the $m_*=225$ $GeV$ for $L=200$ $fb^{-1}$  and $f_{\gamma}/{\Lambda}=1/m_*$. This result has been excluded by the HERA experiment \cite{hera} and we do not need to show results for $L=100$ $fb^{-1}$ and $L=50$ $fb^{-1}$. In figure (\ref{fig4}c) is shown 95\% C.L. excluded region for $f_\gamma/\Lambda$ and $m_*$ plane. This excluded area is smaller than other acceptance region for $m_*<1000$ $GeV$ but for $m_*>1000$ $GeV$ wider than the $0.0015< \xi <0.15$ case, almost same the $0.0015< \xi <0.5$ case.

\section{Conclusions}

Before, we state our main conclusion, it will be useful to remind some facts which are somewhat relevant to our conclusions. Exclusive two lepton production events was observed in measurements of CDF collaborations \cite{cdf1,cdf2,cdf3,cdf4,cdf5,cdf6,cdf7}. These events are consistent in both theoretical expectations with $p\bar{p} \to p\ell^{+}\ell^{-}\bar{p}$ through two photon exchange ($\gamma\gamma \to \ell^{+}\ell^{-}$). The advantage of two lepton final state provides a clean environment due to absence of the proton remnants. This implies that the LHC experiments can rely on process to calculate expectations new physics and luminosity measurements \cite{khoze}. If it is used to measure luminosity then it is important to know its sensitivity to new physics for a given acceptance range. Therefore any signal which conflicts with the SM expectations would be a convincing evidence for new physics. For this reason, we have analyzed the potential of the exclusive production $pp \to p\ell^{+}\ell^{-}p$ to extend the bounds on the model parameters of excited leptons. We have investigated sensitivity of $pp \to p\ell^{+}\ell^{-}p$ to excited leptons three forward detector acceptance; $0.0015< \xi < 0.5$, $0.0015< \xi < 0.15$ and $0.1< \xi <0.5$ and three LHC luminosity values; $50$ $fb^{-1}$, $100$ $fb^{-1}$ and $200$ $fb^{-1}$. We have obtained that the $0.0015< \xi < 0.5$, $0.0015< \xi < 0.15$ forward detector acceptance regions can be improved sensitivity to $m_*$ for $f_\gamma/\Lambda=1/m_*$. In table (\ref{tab3}) we recapitulate the limits at HERA, LEP, Tevatron in comparison with our most stringent results for $L=200$ $fb^{-1}$. $0.1< \xi <0.5$  case is the most clean case and mimics an extra $p_t$ cut ($N_{SM}<0.5$ all of $p_t$ region). Also we have shown the excluded region areas for $f_\gamma/\Lambda$ and $m_*$ plane. $0.1< \xi <0.5$ case is least sensitive for $m_*<1000$ $GeV$ but almost same $0.0015< \xi < 0.5$ case for  $m_*>1000$ $GeV$. These excluded regions are better than the cases of the HERA, LEP and Tevatron colliders.

\pagebreak

\pagebreak

\begin{table}
\caption{Sensitivity of $pp\to p\ell^-\ell^+p$ to the mass of the
at 95\% C.L. for excited lepton mass $m_*$ various values of the $N_{SM}$ and
integrated LHC luminosities $200$ $fb^{-1}$, $100$ $fb^{-1}$ and $50$ $fb^{-1}$. Lower bounds of $m_*$
are given in units of GeV.
We impose different cuts on the transverse momentum of
final leptons to achieve different $N_{SM}$ . Forward detector
acceptance is $0.0015<\xi<0.5$ and $f_\gamma/\Lambda=1/m_*$.
\label{tab1}}
\begin{ruledtabular}
\begin{tabular}{ccccc}
$f_\gamma/\Lambda=1/m_*$ \\
$p_{t,min}$ (GeV) &$N_{SM}$ &$L$=200$fb^{-1}$  &$L$=100$fb^{-1}$  &$L$=50$fb^{-1}$\\
\hline
$460$&0&380&335&290\\
$340$&1&418&353&300\\
$300$&2&420&360&305\\
$280$&3&407&352&302\\
$260$&4&415&342&295\\
$255$&5&390&360&292\\
$210$&10&350&300&288\\
\end{tabular}
\end{ruledtabular}
\end{table}

\begin{table}
\caption{Same as the table \ref{tab1} but for $0.0015<\xi<0.15$.
\label{tab2}}
\begin{ruledtabular}
\begin{tabular}{ccccc}
$f_\gamma/\Lambda=1/m_*$ \\
$p_{t,min}$ (GeV) &$N_{SM}$ &$L$=200$fb^{-1}$ &$L$=100$fb^{-1}$  &$L$=50$fb^{-1}$ \\
\hline
$420$&0&325&292&253 \\
$320$&1&380&328&280\\
$290$&2&375&326&268\\
$270$&3&370&308&278 \\
$250$&4&380&327&261 \\
$240$&5&375&300&260 \\
$200$&10&378&312&267 \\
\end{tabular}
\end{ruledtabular}
\end{table}

\begin{table}
\caption{The obtained limits at HERA, LEP and Tevatron with their references and most stringent results for $L=200$ $fb^{-1}$ and two dedector acceptance region; $0.0015<\xi<0.5$, $0.0015<\xi<0.5$. Lower bounds of $m_*$
are given in units of GeV.
\label{tab3}}
\begin{ruledtabular}
\begin{tabular}{ccccc}
$f_\gamma/\Lambda=1/m_*$ \\
HERA &LEP &Tevatron &LHC $(0.0015<\xi<0.5)$  &LHC $(0.0015<\xi<0.15)$ \\
\hline
$228-272$           &$103-295$                       &$221$        &$420$&$380$ \\
\cite{hera1,hera}   &\cite{lep1,lep2,lep3,lep4}  &\cite{tevatron}  \\
\end{tabular}
\end{ruledtabular}
\end{table}

\pagebreak

\end{document}